11/1/13

# Effect of resonant magnetic perturbations on ELMs in connected double null plasmas in MAST

A. Kirk, Yueqiang Liu, I.T. Chapman, J. Harrison, E. Nardon[1], R. Scannell, A.J. Thornton and the MAST team

*EURATOM/CCFE Fusion Association, Culham Science Centre, Abingdon, Oxon OX14 3DB, UK.*
*[1]Association Euratom/CEA, CEA Cadarache, F-13108, St. Paul-lez-Durance, France*

## Abstract

The application of resonant magnetic perturbations (RMPs) with a toroidal mode number of n=3 to connected double null plasmas in the MAST tokamak produces up to a factor of 9 increase in Edge Localized Mode (ELM) frequency and reduction in plasma energy loss associated with type-I ELMs. A threshold current for ELM mitigation is observed above which the ELM frequency increases approximately linearly with current in the coils. The effect of the RMPs is found to be scenario dependent. In one scenario the mitigation is only due to a large density pump out event and if the density is recovered by gas puffing a return to type I ELMs is observed. In another scenario sustained ELM mitigation can be achieved irrespective of the amount of fuelling. Despite a large scan of parameters complete ELM suppression has not been achieved. The results have been compared to modelling performed using either the vacuum approximation or including the plasma response. The requirement for a resonant condition, that is an optimum alignment of the perturbation with the plasma, has been confirmed by performing a scan in the pitch angle of the applied field.

## *1. Introduction*

The control of Edge Localised Modes (ELMs) during high confinement (H-mode) operation of future fusion tokamaks is essential in order to ensure an acceptable lifetime for plasma facing components [1]. Amongst the various methods considered, the application of non-axisymmetric magnetic perturbations has been intensively studied in a range of devices (for example DIII-D [2][3] JET [4] ASDEX Upgrade [5] and MAST [6]). Most of these studies have been performed in Single Null Diverted (SND) plasma configurations where this technique of Resonant Magnetic Perturbations (RMPs) has been employed to either produce more frequent smaller ELMs (ELM mitigation) or ELM suppression. However, far fewer studies have been performed on Connected Double Null (CDN) discharges. In DIII-D, although it is possible to suppress ELMs in an ITER similar SND shaped plasma ELM suppression could not be attained in a CDN shape but ELM mitigation was achieved [7]. The only experiment to report ELM suppression in a CDN plasma is KSTAR, where suppression was achieved by applying an n=1 RMP (where n is the toroidal mode number) [8].

In this paper the results of applying RMPs in an n=3 configuration to two MAST CDN scenarios will be presented. The parameters of these so called scenario 4 and 6 H-mode plasmas, which are routinely used in MAST are given in sections 2 and 3 respectively. The poloidal cross section, together with the location of the internal coils, for these scenarios is shown in Figure 1. The MAST ELM control system consists of 18 coils in two rows with 6 in the upper row and 12 in the lower row. These coils give considerable enhanced flexibility since they allow improved alignment of the magnetic perturbations with the plasma equilibrium. In addition to the n=3 configurations called "even" (where the currents in the upper and lower coil at the same toroidal location have the same sign) or "odd" parity (the currents in the upper and lower coil have opposite sign), 30L and 90L degree configurations are also possible i.e. where the current in the upper coil has the same sign as a lower coil at a toroidal location displaced by 30 or 90 degrees in the toroidal direction. The effects of the alignment of the perturbation with the plasma equilibrium are normally studied by performing either a scan in plasma current or toroidal field (a so called q scan, where q is the safety factor). However, as will be described in section 4, by using all 18 coils it is possible to vary the ratio of the current in neighbouring coils in the lower row to change the angle of the applied perturbation while keeping the plasma equilibrium constant.

## 2. Effect of RMPs on a scenario 4 discharge

The scenario 4 discharge is based around a neutral beam heated CDN discharge with $I_P = 750$ kA, $B_T = 0.55$ T, $\beta_N \sim 3.8$ (where $\beta_N$ is the normalised plasma pressure) and $q_{95} = 5.4$, which has a larger radius than is usual for MAST discharges. The ERGOS code (vacuum magnetic modelling) [9] has been used to calculate the magnetic perturbations to the plasma due to the application of RMPs. Its implementation on MAST has been previously described in reference [10]. The calculations are performed by adding the vacuum perturbations from the RMP coils to the toroidally symmetric equilibrium reconstructed by EFIT [11]. The equilibrium reconstructions have been constrained using information on the measured pressure profiles and, where available, the pitch angle from a motional Stark effect diagnostic. Previous studies have shown that the uncertainties in the equilibrium have a small effect on the derived parameters [12].

ERGOS modelling shows that this equilibrium was neither aligned with an even or odd parity configuration of the RMPs, the optimum configuration falling approximately half way between the two cases [6]. However, as $q_{95}$ is lowered the modelling shows that in an odd parity configuration of the RMPs the magnetic perturbations become more aligned with the plasma q profile. Previous experiments showed that it was possible to increase the ELM frequency and decrease the ELM energy during the application of RMPs by reducing the $q_{95}$ of the plasma in line with the ERGOS calculations [6]. In the shot with $q_{95} = 5.4$, the n=3 RMPs in an odd parity configuration had no effect, whilst for the shot with $q_{95} = 4.9$ a drop in density was observed followed by an increase in ELM frequency [6].

The installation of the 6 lower coils in MAST means that rather than adjusting the plasma equilibrium to match the applied perturbation it is possible to choose a configuration of the coils which best matches the equilibrium. In the case of the baseline scenario 4 plasma this is a 90L configuration of the coils (one in which the current in the upper coil has the same sign as a lower coil at a toroidal location displaced by 90 degrees in the toroidal direction). Figure 2 shows a plot of the poloidal magnetic spectra of the applied perturbation ($b^{l}$) as a function of poloidal mode number (m) and normalised radius ($\Psi_{pol}^{\frac{1}{2}}$). $b^{l}$ represents the normalised component of the perturbed field perpendicular to equilibrium flux surfaces and is given by $b^1 \equiv \left( \vec{B} \cdot \vec{\nabla} \psi_{pol}^{\frac{1}{2}} \right) \Big/ \left( \vec{B} \cdot \vec{\nabla} \varphi \right)$ where $\vec{B}$ is the total field vector and $\varphi$ is the toroidal angle [13]. Superimposed on the spectra are the locations of the q=m/3 rational surfaces. The applied perturbation is resonant (i.e. the rational surface locations are well aligned with the peaks in the applied perturbation) for this configuration of the coils. The maximum value of the normalised radial field component ($b^r_{res}$) for these shots is $0.65 \times 10^{-3}$.

Figure 3c shows the target $D_\alpha$ time trace for the baseline shot that does not have the RMPs applied, which has a type I ELM frequency of $\sim 100$ Hz and an approximately constant line average density (Figure 3b). This baseline scenario has no gas puffing from 180ms and the refuelling is coming from recycling from the targets and the residual neutral density in the vessel. Figure 3d shows the $D_\alpha$ trace where a rapid burst of ELMs at 265 ms results in a density pump out (Figure 3b) shortly after the current in the ELM coils reaches it maximum value of 5.6kAt. Following the reduction of the density the plasma attains a stable state with high frequency ($\sim 4000$ Hz), small ELMs. The density and electron temperature profiles before, during and after the density pump out event have been measured by a Thomson scattering system (Figure 4). During the density pump out there is a drop in the edge density but little effect on the electron temperature across the whole profile or in the core density. In the rapid ELM-ing phase the density pedestal has reduced significantly from $3.8 \times 10^{19}$ $m^{-3}$ to $1.3 \times 10^{19}$ $m^{-3}$, while the pedestal electron temperature has remained constant at $\sim 150$ eV, the core density and temperature actually rise during this period but the plasma stored energy decreases from 110 kJ to 90 kJ. The pedestal collisionality ($\nu_e^*$) drops from $\nu_e^* \sim 2$ in the type I ELM period to $\nu_e^* \sim 1$ in the rapid ELM-ing phase.

In order to try to recover the loss in density a gas puff of $3 \times 10^{21}$ $D_2 s^{-1}$ has been applied such that it should start to refuel the plasma from ~ 300 ms onwards. The application of this gas puff produces a transition from the high frequency ELMs to a regular low frequency type I ELM-ing regime (Figure 3e) and a recovery in the density (Figure 3b), which is mainly due to a recovery of the density pedestal (Figure 4a), from ~ 340 ms onwards. If this gas puff is applied continually throughout the H-mode period the density pump out and transition to small high frequency ELMs can be removed altogether (Figure 3e) i.e. it appears only to be the reduction in density that is responsible for the smaller high frequency ELMs. It should be noted that in the refuelled case there are a number of double or compound ELMs (at 282 and 290ms for example), which are not observed in the shot without RMPs. On DIII-D, in lower single null discharges, the application of a continuous deuterium gas puff in the divertor region to a previously ELM suppressed discharge leads to small grassy ELMs appearing as the pedestal density is increased [14]. At higher gas puff rates large compound type I ELMs appear that have similar characteristics to those observed in Figure 3e. For the DIII-D case the result is explained as being due to a change in collisionality, with ELM suppression only being established for $v_e^* < 0.2$ [14]. In these MAST discharges $v_e^* > 1.0$, which may explain why full suppression is not observed.

In order to investigate if the small ELM period observed in MAST is just a consequence of the reduction in density, a shot has been designed in which the initial fuelling is reduced to such a point that the plasma just has sufficient density to attain H-mode at the input power used (3.4 MW). Figure 5c shows the $D_\alpha$ time trace for this low density shot. After the transition to H-mode at 230 ms there are a few large ELMs followed by a period (from 250 to 280 ms) of small high frequency ELMs. Although the ELM frequency is not as high as in the case where the RMPs are applied (2.5 kHz compared to 4 kHz) the characteristics are sufficiently similar to confirm that these small ELMs are due to the reduction in density. As the density increases (from 280 ms onwards) larger ELMs reappear. Previous studies of this low density discharge [15] have identified the small ELMs as being type IV in nature (i.e. the low collisionality branch of type III ELMs) and a peeling ballooning stability analysis revealed that the edge parameters associated with these ELMs lie in a completely ideal MHD stable region, suggesting that they are associated with a resistive MHD stability [15].

The density pump out event appears to be associated with the triggering of a burst of high frequency ELMs but it is not clear if this is cause or effect. Figure 6 shows the radial profiles of the lower divertor target heat flux measured by infrared thermography at four times during shot 26848. Figure 6a shows the profile at 0.23 s in an inter-ELM period just before the RMPs are applied. The profile is consistent with being exponential in the scrape off layer region. The profile in Figure 6b is obtained in an inter-ELM period after the RMPs have reached their maximum value but before the density pump out event has occurred. At this time two sub peaks can be observed at ~ 1.12 and 1.18 m. The splitting of the divertor strike point is a common observation in the presence of external non-axisymmetric magnetic perturbations (see for example [16][17][18]) and is of particular interest because it may give indications on the degree of magnetic stochasticity at the edge of the plasma and thereby help to understand the physical mechanisms at play.

Superimposed on this figure as a dashed curve is the "radial field line excursion profile", calculated by field line tracing. The field lines are traced from the target region and the deepest radius that they reach, in terms of the unperturbed $\psi^{1/2}_{pol}$ is calculated. The "radial field line excursion' profile, is the profile of $1 - (\psi^{1/2}_{pol})^{min}$ calculated at the toroidal angle location of the infrared diagnostic. Although the relationship between the radial excursion and deposited heat flux is not trivial, it might be expected that field lines having a larger excursion would carry a larger heat flux (see e.g. [19]). The reasonably good correspondence in the position of the peaks in the heat flux and radial excursion profiles suggests that the strike point splitting may be due to a change in the magnetic field structure imposed by the n=3 RMPs. In contrast, at the start of the density pump out event at 266 ms, Figure 6c, a large peak is observed at ~ 1.08 m. This profile with a secondary peak larger than the primary is typically observed in MAST during the presence of a locked mode. The locked-mode like profile lasts for ~ 2ms before returning to a more typical profile as shown in Figure 6d. In this profile it is difficult to observe any splitting due to the RMPs, this could be because the splitting is absent or it is not visible due to the high frequency (~ 4 kHz) of the ELMs and the integration time (200 μs) of the detector.

Locked modes are normally identified on MAST using saddle coils which are located on the vessel wall but due to the characteristics of the wall a locked mode lasting less than 2 ms would not be detected. Similarly there is no evidence for a change in the plasma rotation measured using the charge exchange system, but since this measurement integrates over a 5ms period it may also not detect a short lasting locked mode. Further experiments will be performed in future campaigns to try to identify the cause of the density pump out but since the ELMs are only mitigated while the density is reduced it would seem that this is not a valid ELM mitigation scenario to pursue.

### 3. *Sustained ELM mitigation in a scenario 6 discharge with RMPs in an n=3 even parity configuration*

The scenario 6 discharge is based around a neutral beam heated CDN discharge with $I_P$ = 550 kA, $B_T$ = 0.585 T, $\beta_N\sim$ 4.0, $q_{95}$= 6.8 and an outer plasma radius of 1.44 m. This shot is continually fuelled using a high field side (HFS) mid-plane gas puff. Figure 7c shows the target $D_\alpha$ time trace for the baseline shot that does not have the RMPs applied, which has a type I ELM frequency of ~ 70 Hz and a slowly rising line average density (Figure 7b). Figure 8a shows that vacuum modelling predicts that if the RMPs are applied in an even parity configuration then the rational surface locations are well aligned with the peaks in the applied perturbation. As can be seen from Figure 7d the application of the RMPs with $I_{ELM}$ = 5.6 kAt in an even parity configuration produces an increase in ELM frequency to $f_{ELM}$ ~ 230 Hz (i.e. a factor of ~3.5 increase over the natural ELM frequency). There is a small reduction in density but unlike in the scenario 4 discharges discussed in the last section, no large pump out event is observed. In the shot without RMPs the stored energy ($\beta_N$) rises from 60 kJ (3.0) at 0.18 s to 90 (4.2) at 0.24 s at which point a tearing mode is destabilised in the plasma which first clamps the rise in stored energy and then causes a back transition to L-mode at 0.27 s. In the shot with RMPs applied the stored energy reaches a flat top value of 82 kJ with $\beta_N$ = 3.9 and the tearing mode is not triggered, which allows the discharge to survive and the ELM mitigation to continue until either the ELM coil current is reduced or the end of the solenoid swing is reached and the plasma current is ramped down at 0.42 s.

Repeat discharges have been performed with increasing current in the coils ($I_{ELM}$) to determine the threshold current for the onset of ELM mitigation together with the effect on ELM frequency. For shots with $I_{ELM} < 4.0$ kAt no increase in ELM frequency is observed. Figure 9a shows that the mitigated ELM frequency ($f_{mit}$) as a fraction of the natural ELM frequency ($f_{natural}$) rises linearly as a function of $I_{ELM}$ above the threshold value.

Another way of varying the size of the applied RMP is to vary the distance of the plasma from the coils. The baseline scenario has an outer radius or 1.44 m, which means that the distance between the coils and plasma edge is 28 cm. Repeat shots have been performed with $I_{ELM} = 5.6$ kAt where the outboard radius of the plasma was increased up to 1.50 m. The ELM frequency relative to the frequency in a reference shot at the same radius with no RMPs increases as the distance to the coils decreases (see Figure 9b). The ELM frequency relative to the RMP off shot increased by up to a factor of 8.5. It was not possible to increase the radius beyond $R_{OUT} = 1.50$ m due to large interactions of the plasma with in vessel components and even at this radius minor interactions could be observed, especially during ELMs. Therefore the remaining scans were performed at $R_{OUT} = 1.48$ m, which resulted in an increase in ELM frequency over the RMP off case of a factor of 6.5.

In order to test the sensitivity of the ELM frequency to the alignment of the applied perturbation with the pitch of the equilibrium magnetic field a scan in $q_{95}$ has been performed. The vacuum modelling calculations shown in Figure 8b suggest that if the $q_{95}$ was increased to ~7.6 the alignment, especially towards the edge region of the plasma, could be improved. Since this scenario is already using the maximum toroidal field the only way to increase the edge safety factor is to reduce the plasma current. Figure 10c shows the target $D_\alpha$ time trace for a shot with $I_P$ = 500 kA that does not have the RMPs applied, which has an ELM frequency of ~ 125 Hz. The outer radius of the plasma is $R_{OUT}$ =1.48 m, and $q_{95}$ = 7.6. The peak stored energy in this shot is reduced to 70 kJ and the lower value of $\beta_N$ = 3.8 means that a tearing mode is not triggered and the shot continues after 0.3 s. The application of the RMPs with $I_{ELM}$ = 5.6 kAt in an even parity configuration, Figure 10d, produces an increase in ELM frequency to $f_{ELM}$ ~ 930 Hz (i.e. a factor of ~7.5 increase over the natural ELM frequency) and a reduction in stored energy to 60 kJ and $\beta_N$ =3.2. Figure 9c shows the mitigated ELM frequency ($f_{mit}$) as a fraction of the natural ELM frequency ($f_{natural}$) as a function of $q_{95}$ for plasmas with $R_{OUT}$ =1.48 m, which shows that for the range of $q_{95}$ explored (6.8 – 7.6) ELM mitigation can be established and the largest increase in ELM frequency was obtained for $q_{95}$=7.6.

Scans in line average density were also performed but the optimum density, which is the one that produced the largest increase in ELM frequency, was the one shown in Figure 10.

### 3.3 Vacuum and plasma response modelling of the increase in ELM frequency

In order to investigate if the changes in ELM frequency observed can be correlated with parameters determined from vacuum modelling the ERGOS code [9] has been used to calculate the maximum value of the normalised radial field component ($b^r_{res}$) for all the shots considered. Figure 11 shows that in all cases, ELM mitigation is observed when $b^r_{res}$(thresh) > $0.5x10^{-3}$. There is a significant difference in the dependence of the normalised ELM frequency with $b^r_{res}$ for the three scans suggesting that the value of the field at a single location is not sufficient and that instead some measure of the radial dependence of the perturbations, including a description of penetration of the field is required. However, the scans in $I_{ELM}$ and $R_{OUT}$ do point to a similar threshold value of $b^r_{res}(thresh) \sim 0.5 \times 10^{-3}$, which is similar to what was found in a previous study of lower single null discharges in MAST where the threshold for ELM mitigation was found to be $b^r_{res}(thresh) \sim 0.4 \times 10^{-3}$ for RMPs in an n=4 configuration and $b^r_{res}(thresh) \sim 0.55 \times 10^{-3}$ in an n=6 configuration [20]. However, since a threshold in $q_{95}$ was not observed a similar conclusion can not be drawn from that data.

Calculations have been performed using the MARS-F code, which is a linear single fluid resistive MHD code that combines the plasma response with the vacuum perturbations, including screening effects due to toroidal rotation [21]. In this case the resistive plasma response reduces the resonant component of the field near the edge of the plasma ($\Psi_{pol}=0.98$) by a factor of 4 relative to the vacuum approximation. The MARS-F calculations have been performed for the discharges used in the $I_{ELM}$ scan and the value of $b^r_{res}$ taking into account the plasma response has been calculated. Figure 12a shows the normalised ELM frequency versus $b^r_{res}$ with the plasma response included. The threshold value $b^r_{res}(thresh) \sim 0.14 \times 10^{-3}$ is again similar to the threshold value obtained in the LSND experiments [20] where the threshold value was $0.10 \times 10^{-3}$ and $0.15 \times 10^{-3}$ for the RMPs in an n=4 and n=6 configuration respectively.

In the MARS-F modelling the RMP field causes a 3D distortion of the plasma surface (Figure 12b), which potentially leads to the formation of a 3D steady state equilibrium. This plasma displacement varies with toroidal angle ($\phi$) as $\xi e^{in\phi}$ where n is the toroidal mode number and $\xi$ is the amplitude of the normal displacement of the plasma surface which varies as a function of poloidal angle ($\theta$). Previous MARS-F simulations of the effect of RMPs on the MAST plasma showed a clear correlation between the location of the maximum of the amplitude of the normal component of the plasma displacement at the plasma surface and the effect of the RMPs on the plasma [22]. In these studies it was observed that a density pump out in L-mode or ELM mitigation in H-mode only occurred when the displacement at the X-point was larger than the displacement at the mid-plane. Figure 12b shows this displacement ($\xi$) as a function of $\theta$, calculated at the $\Psi_{pol}$= 0.98. The displacement has a peak near the X-points ($\theta \sim -90°$), with a maximum displacement of the order of 2.1 mm/kAt i.e. for the coil current used (5.6kAt) the maximum displacement is $\sim$ 12 mm. If the RMPs are applied in an odd parity configuration then no change in the ELM frequency is observed experimentally and, similar to what was reported in [6], the MARS-F code calculates that the displacement ($\xi$) is peaked at the mid-plane. This again suggests that it is the location of the peaking rather than just the size of the displacement that is important.

Figure 12c shows the normalised ELM frequency versus displacement at the X-point, the threshold value obtained is $\xi > 7$mm which is much larger than the threshold displacement of $\sim$ 1.5 mm determined from the LSND discharges. So while it is generally true that a clear correlation exists between the location of the maximum and the effect of the RMPs on the plasma, unlike in the case of $b^r_{res}$(thresh), there does not seem to be a unique value for the amplitude of this displacement. Although for a given equilibria $b^r_{res}$ and $\xi$ scale linearly with the current in the RMPs, there is not a simple analytical scaling between the two quantities. This is because the magnitude of the plasma displacement depends on both the resonant and non-resonant components of the applied field and the poloidal location of the peak in the displacement is due to what response is excited in the plasmas (either peeling-tearing or kink-like) [22], which depends on the underlying plasma equilibrium.

The toroidal rotation velocity ($V_\phi$) has been measured in these Scenario 6 discharges as a function of time using a charge exchange recombination spectroscopy system. Figure 13 shows the radial profiles of $V_\phi$ as a function of time for the shots shown in Figure 7. There is a slight modification of the rotation profile near the edge of the plasma (R= 1.2-1.3m) at a time of 190 ms during the time the RMPs are being increased but thereafter the velocity profiles are very similar.

This is in sharp contrast to what was observed when the RMPs were applied in an n=3 configuration to the LSND discharges [20], where a strong braking of the toroidal rotation was observed. This braking was so severe that it produced a back transition to L-mode before any sustained ELM mitigation could be achieved. The quasi-linear MARS-Q code [23] has been used to simulate the RMP penetration dynamics and the toroidal rotation braking for both the LSND and CDN shots. As was reported in reference [20] the simulations showed that a full damping of the toroidal rotation, which is initially due to the $\vec{j} \times \vec{B}$ torque, is achieved in a time of less than 40ms, very similar to what is observed in the experiment. For the scenario 6 discharges presented here the toroidal rotation velocity ($V_\phi$) has been measured using charge exchange recombination spectroscopy. The symbols in Figure 14 show the core value of $V_\phi$ (measured at normalised poloidal flux $\Psi_{pol}$ =0.3) as a function of time after the RMPs have reached their maximum value. Similar to what is observed experimentally the the MARS-Q code also predicts little braking in this plasma configuration (see the line in Figure 14).

A possible explanation for this difference may lie in the fact that the CDN plasmas have a much larger $q_{95}$ ($q_{95} \sim 7.0$ compared to 3.0 in the LSND), which means that there are more rational surfaces near the edge of the plasma which can screen the perturbation and hence reduce the $\vec{j} \times \vec{B}$ torque on the core of the plasma. These differences will be explored further in future modelling and experimental studies.

### 3.2 Plasma parameter range explored and effect on ELM size

While clear ELM mitigation has been observed in these shots, ELM suppression has not been established despite a large range of plasma parameters being scanned. On DIII-D ELM suppression has been observed when the pedestal collisionality ($v_e^*$) is < 0.3 [3] or > 2.0 [2]. While on ASDEX Upgrade the suppression of type I ELMs is associated with a plasma density expressed as a fraction of the Greenwald density ($n_{GW}$), with suppression being observed for $n_e/n_{GW}>0.53$ [24]. For the MAST CDN discharges considered here the collisionality range scanned is shown in Figure 15a and the density as a fraction of the Greenwald density is shown in Figure 15b. The data overlaps with both the high collisionality region for DIII-D and the $n_e/n_{GW}$ region required for suppression on ASDEX Upgrade. However, the ranges stated above are for SND plasmas on both devices and suppression in CDN discharges has not been reported on either. In fact, on DIII-D ELM suppression was not achieved in low collisionality CDN discharges [7]. The only machine to report ELM suppression in CDN discharges is KSTAR [8], which used RMPs in an n=1 configuration to suppress ELMs in discharges with collisionalities in the range 0.5-1.0. Hence there does not appear to be a unique range that ensures ELM suppression in CDN discharges.

Turning to the effect that the increased ELM frequency has on the ELM size and peak heat fluxes to the target. Figure 16a shows a plot of the energy loss per ELM ($\Delta W_{ELM}$), derived from the change in plasma stored energy calculated by the EFIT equilibrium code [11], versus $f_{ELM}$ for the natural and mitigated ELMs. The application of the RMPs produces an increase in $f_{ELM}$ and corresponding decrease in $\Delta W_{ELM}$ consistent with $f_{ELM}.\Delta W_{ELM}$= const (represented by the solid curve in Figure 16a). Not only does the total energy lost per ELM decrease but so does the number of particles. The change in plasma density due to an ELM expressed as fraction of the pre-ELM pedestal density has a mean value of $\Delta n_e/n_e^{ped} = 0.04$ for the natural ELMs, which decreases to 0.02 at the highest ELM frequencies (Figure 16b).

In order to avoid damage to in-vessel components in future devices, such as ITER, it is the peak heat flux density at the divertor that is important rather than the actual ELM size. The divertor heat fluxes on MAST have been measured using infrared thermography. Figure 16c shows the peak heat flux density at the target ($q_{peak}$) as a function of $\Delta W_{ELM}$. The increase in ELM frequency and decrease in $\Delta W_{ELM}$ does lead to reduced heat fluxes at the target, although it also results in a smaller wetted area at the target meaning that the reduction in $q_{peak}$ is not the same as the reduction in $\Delta W_{ELM}$. The mitigated and natural ELMs follow the same trend and show that a reduction of a factor of 3 in $\Delta W_{ELM}$ (i.e. from 6 to 2kJ) produces a reduction in $q_{peak}$ of 2.6 (from 9 to 4.3 MWm$^{-2}$). A similar weaker than linear decrease in the divertor hear flux with increasing ELM frequency has also been observed during mitigated ELMs on JET [18].  In order to make extrapolations over a wider range it will be important to understand what happens at very small energies, since the extrapolation based on the present data would indicate a peak heat flux of $\sim 2$ MWm$^{-2}$ for $\Delta W_{ELM} = 0$, to be compared to the typical inter-ELM peak heat fluxes of $\sim 0.5$ MWm$^{-2}$.

### 3.3 Effect of the RMPs on the pedestal and ELM stability

The pedestal electron density and temperature characteristics have been measured using a Nd YAG Thomson Scattering (TS) system.  The application of the RMPs in an n=3 even parity configuration causes a toroidally local outward movement of the plasma edge of ~10 mm with respect to the RMP off case.  The pedestal evolution in the inter-ELM period is similar in both the RMP on and off case with both the pedestal density and width increasing during the inter-ELM period [25].  The difference is that in the RMP applied case this evolution is terminated sooner due to the increase in ELM frequency.  Throughout the inter-ELM period the pedestal width is also larger in the RMP applied case, which results in a lower pedestal pressure gradient.

The radial pedestal profiles for a shot with $I_P$=500 kA without RMPs and with RMPs in an n=3 even parity configuration, both obtained in the last 10 % of the ELM cycle, are shown in Figure 17a, b and c for the electron density, temperature and pressure gradient respectively mapped onto normalised flux, using the unperturbed equilibrium.  In radial space a shift of the pedestal profile is observed of the order of ~1cm due to the application of the RMPs.  To compensate for this displacement when mapping to poloidal flux, the profiles have been aligned using a constraint based on the power crossing the Last Closed Flux Surface (LCFS) that sets the electron temperature at the LCFS to be ~40 eV. A clear drop is observed in the pedestal density but little change in the electron temperature and the peak pedestal pressure gradient reduces.

A stability analysis has been performed on these discharges using the ELITE stability code [26]. The procedure used to analyse the edge stability in MAST has been described in [27]. It consists of reconstructing the equilibrium using the kinetic profiles obtained from the TS system as constraints and assuming that $T_i = T_e$. The edge pressure gradient is then varied at a fixed current density and the edge stability evaluated using ELITE [26].

Figure 18 shows the stability boundary and the experimental point in a plot of peak edge current density ($j_\phi$) versus normalised pressure gradient ($\alpha$) for the discharge without RMPs and for the discharge with the RMPs in an n=3 even parity configuration. The results show that for the discharge without RMPs the experimental point lies on the boundary of the region unstable to peeling-ballooning modes, a trait often associated with type I ELMs. However, for the point with RMPs the analysis predicts that such a discharge would be stable to peeling-ballooning modes and so it is not apparent why the ELM frequency should be higher in such a discharge. ELMs are also triggered in this stable region during application of RMP in SND plasmas in MAST and it has been proposed [28] that the mechanism for degrading the peeling-ballooning stability boundary is the presence of 3D perturbations to the plasma equilibrium.

### *4. Effect of changing the pitch angle of the magnetic perturbation*

The traditional way of trying to identify a resonant window for the effect of RMPs is to keep the applied RMP field fixed while changing the plasma equilibrium, typically by performing a $q_{95}$ scan. This technique has been useful for identifying resonant windows for ELM mitigation on JET [4] and ELM suppression on DIII-D [3] but it has the disadvantage that by performing the scan in $q_{95}$ scan other plasma parameters are changed. In an n=3 configuration MAST can exploit a unique capability allowed by having 12 coils in the lower row. If the current is kept fixed in the 6 upper coils, then by operating the lower coils in pairs and by varying the relative current in the pair the pitch angle of the applied field can be changed. This allows a resonance condition to be studied without changing the underlying plasma parameters.

Table 1 shows the coil currents used on a series of repeat scenario 6 plasmas with a plasma current of 600 kA and an outer radius of 1.47 m. The current in the coil pairs has been arranged such that the combined value summed in quadrature is equal to that in the upper coils. Since all 18 coils have to be powered at some stages in the scan, the voltage limitations of the power supplies mean that the maximum current that can be produced is 1 kA (4 kAt).

The pitch angle ($\alpha$) of the applied perturbation is defined as the angle between the centre of the upper coil and the centroid of the paired lower coils with the same sign. Of course, there are a number of possible angles, for example, in the even parity configuration where coils at the same toroidal location have the same sign in the upper and lower row of coils the angle could be chosen as that between upper coil 1 and lower coils 1 (90°) between upper coil 1 and lower coil 5 (25°) or upper coil 1 and lower coil 9 (13°). We have chosen for the definition of ($\alpha$) the angle that is nearest to the pitch angle of the plasma equilibrium field lines at $\Psi_N$=0.95 at the LFS, which for the equilibrium considered here is ~30°. Hence in the two extreme coil configurations of even parity and 90L, $\alpha$ is defined to be 25° and 32° respectively.

Repeat discharges have been performed with varying pitch angles and the ELM frequency determined (Figure 19a). The current limitation of 1 kA (4kAt) is very close to the threshold value of $I_{ELM}$ = 4.0 kAt determined previously for the onset of ELM mitigation using an even parity configuration of the coils in this discharge. Hence when this configuration is applied to the shot only a small change in ELM frequency over the natural frequency is observed. The ELM frequency is maximised for pitch angles in the range 26.8° to 29.5°.

Vacuum modelling has been performed using the ERGOS code and the maximum value of the normalised radial field component ($b^r_{res}$) calculated as a function of $\alpha$ is shown in Figure 19b. The change in $b^r_{res}$ is correlated with the change in ELM frequency but the size of the change is quite modest over the scan in $\alpha$. This would suggest an onset of mitigation for $b^r_{res}(thresh) > 0.48 \times 10^{-3}$ which is similar to the value obtained in the $I_{ELM}$ scan of $b^r_{res}(thresh) > 0.5 \times 10^{-3}$ reported above.

## 5. Summary and discussion

RMPs in an n=3 configuration have been applied to two type I ELMing H-mode scenarios, which are in a CDN magnetic configuration in MAST. In a low fuelled version of the scenario 4 discharge the RMPs cause a density pump out, which is followed by a small ELM regime. However, the small ELM regime is a consequence of the lower density and refuelling these discharges leads to a return to type I ELMs. The application of the coils to a scenario 6 discharge that is constantly fuelled leads to sustained ELM mitigation, without a dramatic pump out event. The reason for the difference in behaviour between these two discharges is not understood. However, it may be related to the different gas puffing location in the two discharges. The discharge in which ELM mitigation is observed uses a toroidally localised gas puff located at the high field side (HFS) midplane whilst the shot where mitigation is not observed uses a toroidally distributed gas puff located again at the but this time in the X-point region. On DIII-D application of a gas puff in the divertor region also led to a loss of ELM suppression [14]. To test if the location of the gas puff is important for ELM mitigation studies will the performed in the next campaign on MAST by swapping the gas puff location between the two plasma scenarios.

In the scenario 6 discharges the ELM frequency increases by up to a factor of 8.5 with a similar reduction in ELM energy loss. A reduction in ELM size by a factor of 3 leads to a decrease in the peak heat flux by a factor of 2.6. ELM mitigation has been observed for the whole range of shots covered ($0.3 < \nu_e^* < 2.8$, $0.3 < n_e/n_{GW} < 0.8$) but ELM suppression has not been observed. A threshold current for ELM mitigation is observed above which the ELM frequency increases approximately linearly with current in the coils. The ratio of the ELM frequency to the natural frequency also increases strongly as the gap between the plasma and the RMP coils is reduced. In the case of the $I_{ELM}$ and gap scan calculations in the vacuum approximation using the ERGOS code show a linear response above a threshold value of the maximum value of the normalised radial field component ($b^r_{res}$), which indicates that ELM mitigation is observed when $b^r_{res}(thresh) > 0.5 \times 10^{-3}$.

An edge stability analysis shows that the application of the RMPs move the experimental point from a region unstable to peeling-ballooning modes to a stable region. So in such a 2D analysis it is not apparent why the ELM frequency should be higher in the discharges with RMPs. However, during the application of the RMPs the plasma edge takes on 3D distortions, which as was discussed in [28] degrade the peeling-ballooning stability boundary and hence result in more frequent ELMs.

A scan in $q_{95}$ produces a change in ELM frequency in both shots with and without RMPs applied, making the identification of a resonance condition difficult. However, it has been possible to identify such a condition by exploiting the unique ability allowed by the lower 12 ELM coils to perform a scan in the pitch angle of the applied field. The results of this scan show that there is an optimum alignment of the perturbation with the plasma equilibrium field, which maximises the ELM frequency. The experimentally determined optimum alignment is also where vacuum modelling calculations predict the largest value of $b^r_{res}$.

**Acknowledgement**

This work was funded partly by the RCUK Energy Programme under grant EP/I501045 and the European Communities under the contract of Association between EURATOM and CCFE. The views and opinions expressed herein do not necessarily reflect those of the European Commission. This work was carried out within the framework of the European Fusion Development Agreement.

**Tables**

**Table 1 Current in the ELM coils (in kA) used to produce the pitch angle (α) scan.**

| Sector | 1 | 2 | 3 | 4 | 5 | 6 | 7 | 8 | 9 | 10 | 11 | 12 |
|---|---|---|---|---|---|---|---|---|---|---|---|---|
| Upper coil | -1.0 | | +1.0 | | -1.0 | | +1.0 | | -1.0 | | +1.0 | |
| Lower coils | | | | | | | | | | | | |
| α=25° | -1.0 | | +1.0 | | -1.0 | | +1.0 | | -1.0 | | +1.0 | |
| α=26.8° | -0.85 | +0.5 | +0.85 | -0.5 | -0.85 | +0.5 | +0.85 | -0.5 | -0.85 | +0.5 | +0.85 | -0.5 |
| α=28.5° | -0.7 | +0.7 | +0.7 | -0.7 | -0.7 | +0.7 | +0.7 | -0.7 | -0.7 | +0.7 | +0.7 | -0.7 |
| α=29.5° | -0.64 | +0.77 | +0.64 | -0.77 | -0.64 | +0.77 | +0.64 | -0.77 | -0.64 | +0.77 | +0.64 | -0.77 |
| α=30.3° | -0.5 | +0.85 | +0.5 | -0.85 | -0.5 | +0.85 | +0.5 | -0.85 | -0.5 | +0.85 | +0.5 | -0.85 |
| α=31° | -0.3 | +0.95 | +0.3 | -0.95 | -0.3 | +0.95 | +0.3 | -0.95 | -0.3 | +0.95 | +0.3 | -0.95 |
| α=32° | | +1.0 | | -1.0 | | +1.0 | | -1.0 | | +1.0 | | -1.0 |

**Figures**

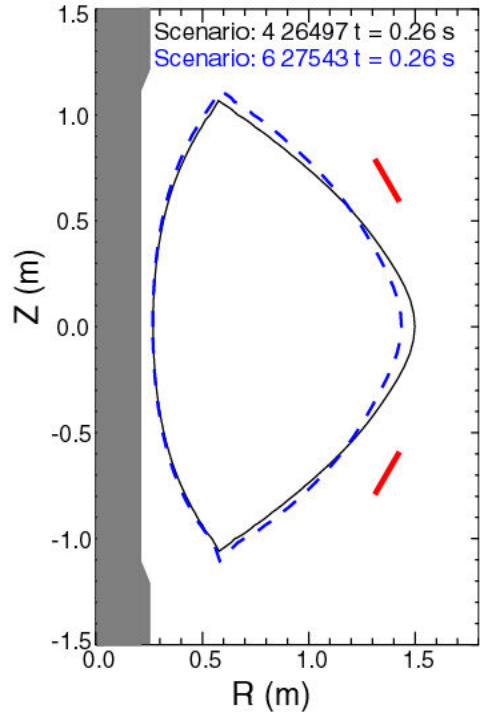

**Figure 1** Poloidal cross section of the scenario 4 (solid curve) and scenario 6 (dashed) plasmas together with the location of the centre column and ELM coils.

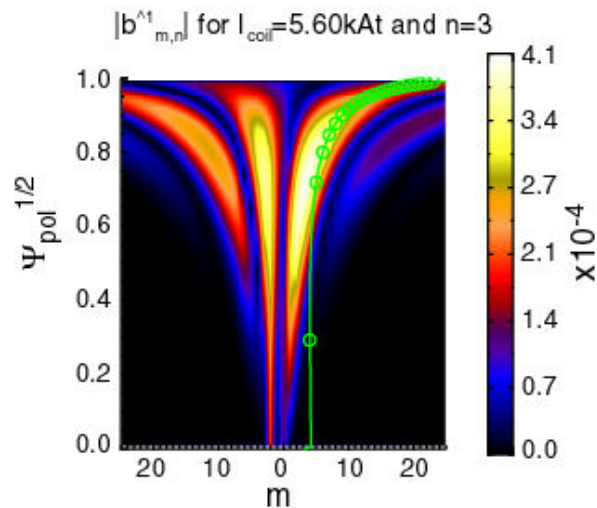

**Figure 2** Poloidal magnetic spectra calculated in the vacuum approximation for a scenario 4 shot with $I_P$=750 kA and a 90L configuration of the coils. Superimposed as circles and line are the q=m/3 rational surfaces of the discharge equilibrium.

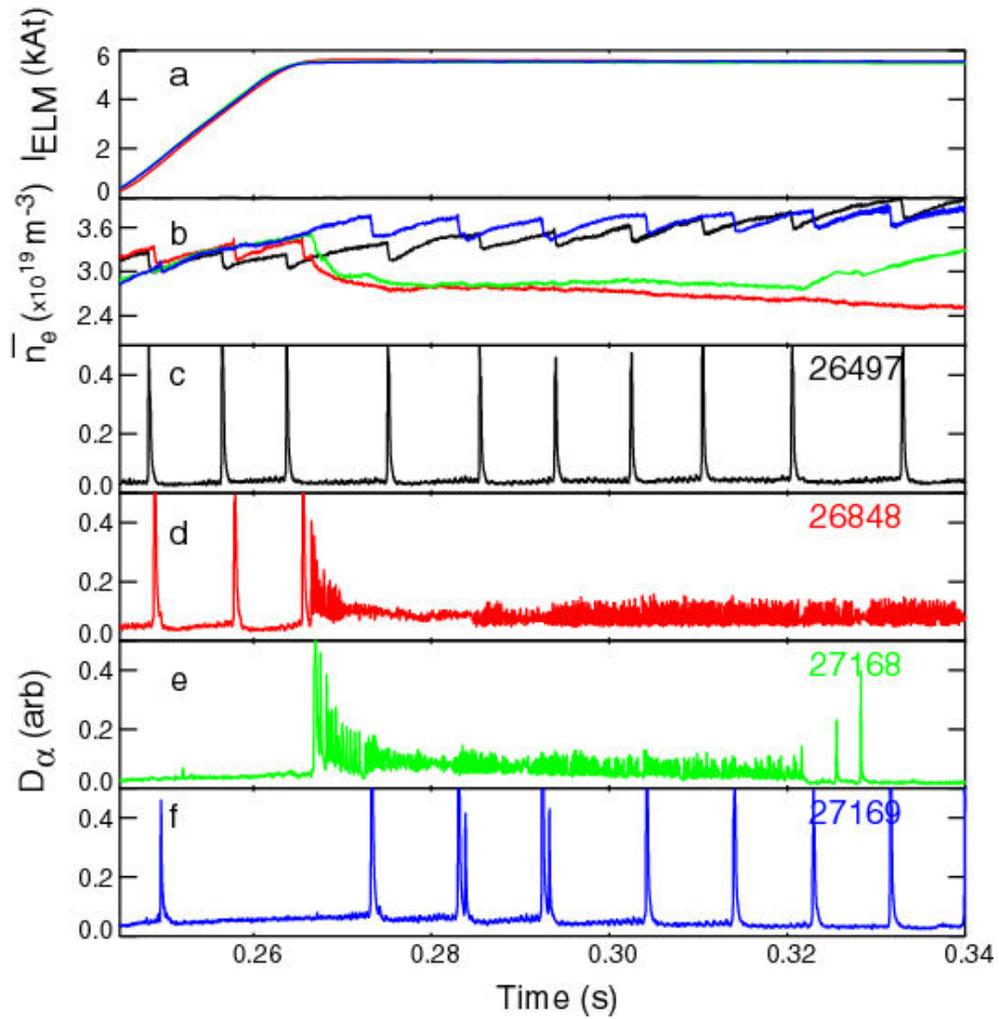

**Figure 3** Time traces for a scenario 4 shot of a) coil current ($I_{ELM}$)  b) line average density ($\bar{n}_e$), and  $D_\alpha$ traces for c) shot  26497 without RMPs and d-f) shots with the RMPs in a 90L configuration with d) no gas puff, e) a gas puff that takes effect from 0.32 s and f) continuous gas puffing.

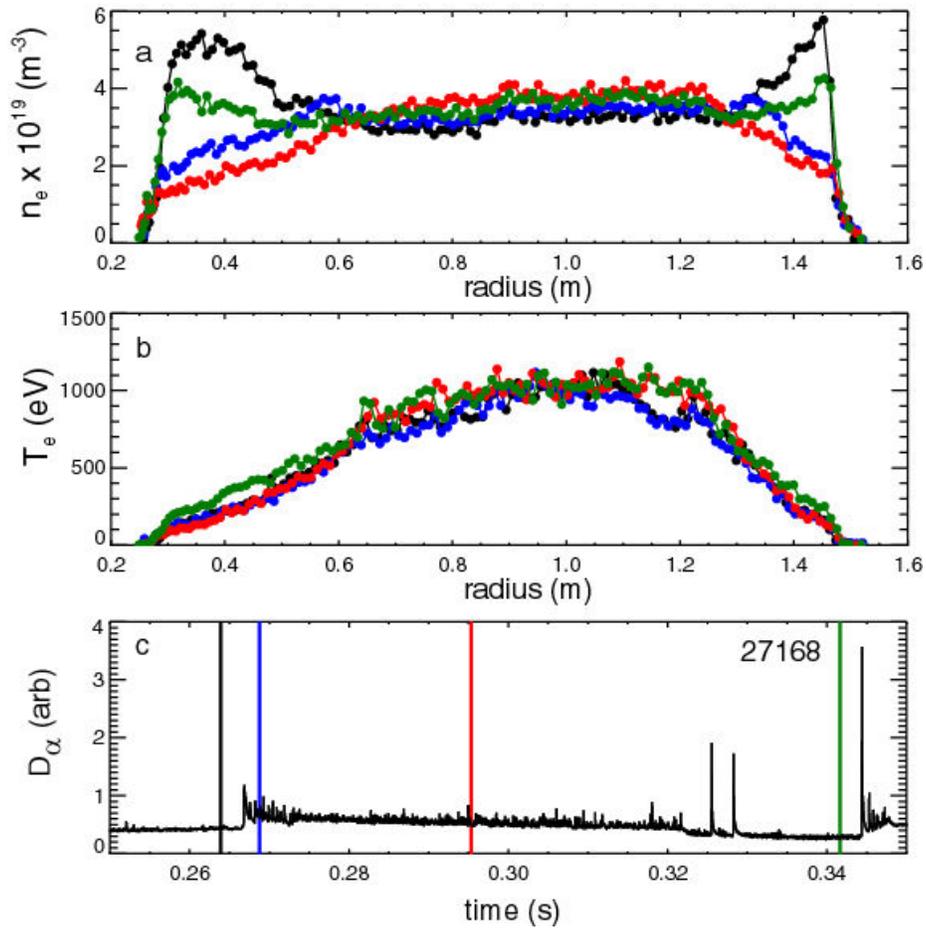

**Figure 4** The radial profiles of a) electron density (n$_e$) and b) the temperature (T$_e$) obtained before the density pump out event (263 ms) during the event (268 ms), in the rapid small ELM period (295 ms) and in the recovered type I ELM period (342 ms) indicated on the c) target D$_\alpha$ time trace.

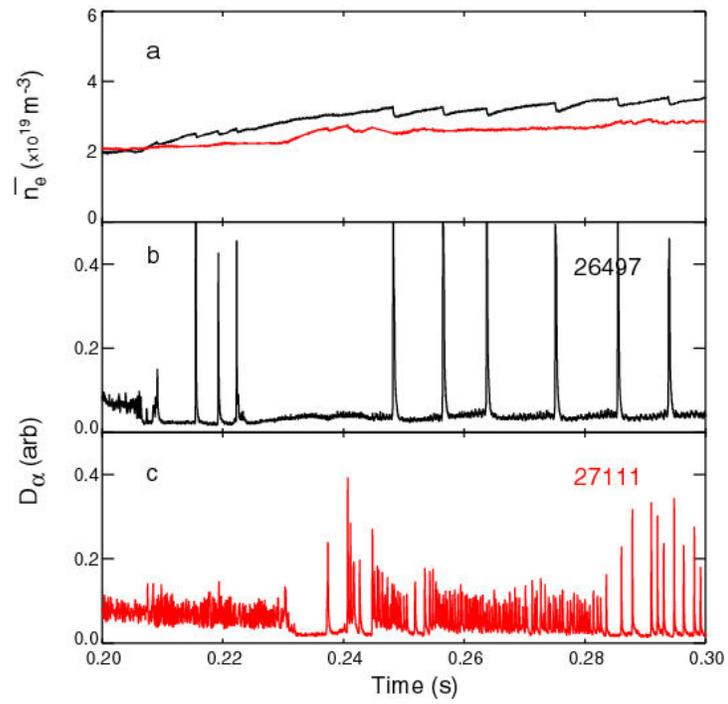

**Figure 5** Time traces of a) line average density ($\bar{n_e}$), and $D_\alpha$ traces in shots without RMPs applied for b) shot 26497 with normal fuelling and c) 2711 with reduced fuelling.

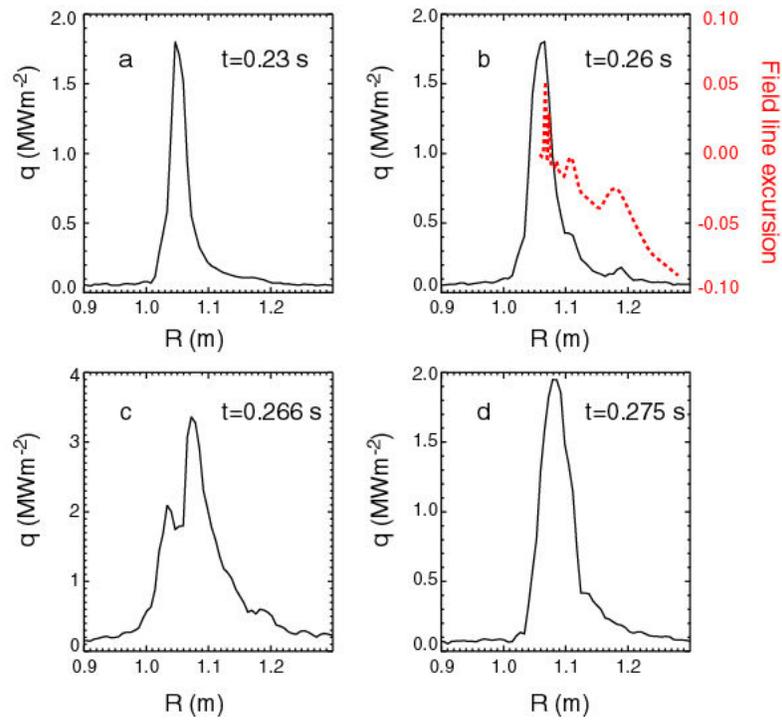

**Figure 6** Radial profiles of feat flux to the lower divertor target measured by infrared thermography during four period of a scenario 4 discharge (26848) in which n=3 RMPs in a 90L configuration. Superimposed on b) as a dotted line is the field line radial excursion (1-$\Psi_{pol}^{1/2}$)$^{min}$, where $\Psi_{pol}^{min}$ is the deepest radius reached by a field line) calculated from vacuum modelling.

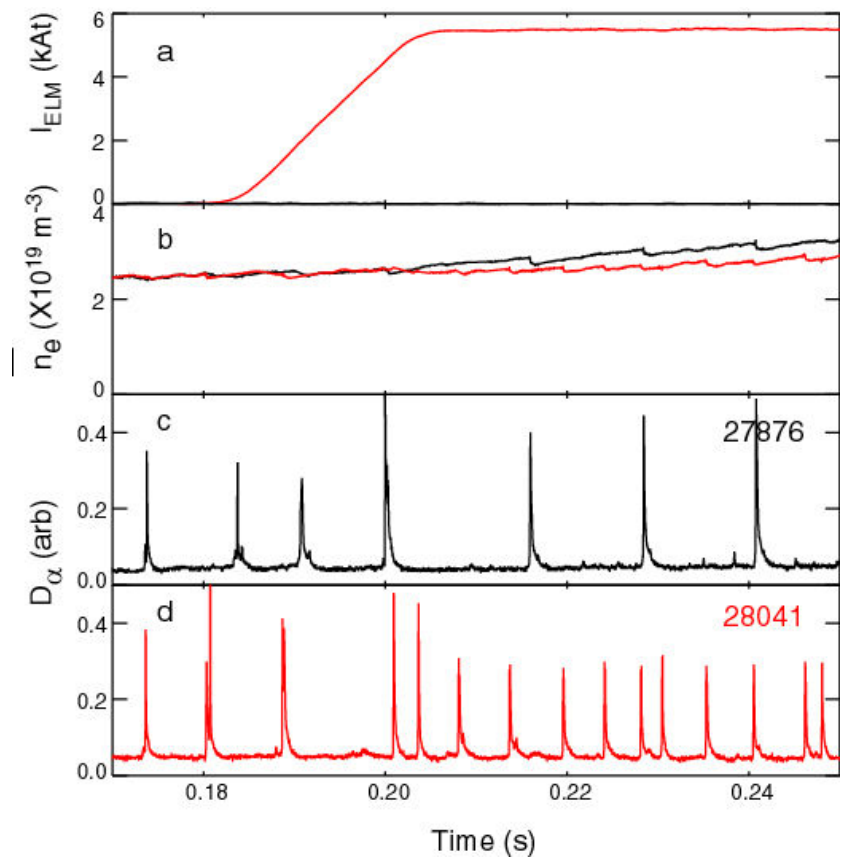

**Figure 7** Time traces for a scenario 6 shot with $I_P=550$ kA of a) coil current ($I_{ELM}$) b) line average density ($\bar{n_e}$), and $D_\alpha$ traces for c) shot 27876 without RMPs and d) shot 28041 with the RMPs in a even parity configuration.

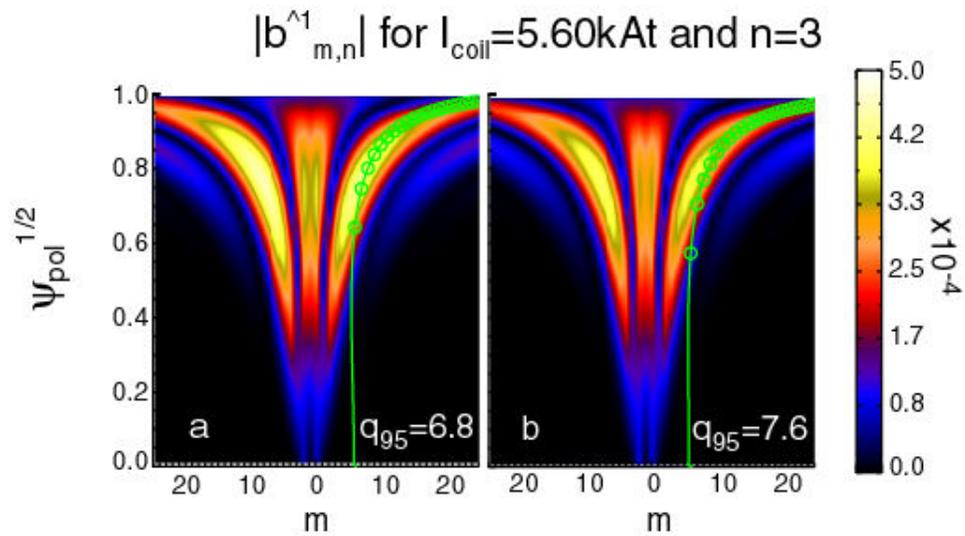

**Figure 8** Poloidal magnetic spectra calculated in the vacuum approximation for RMPs in an even parity configuration applied to a scenario 6 shot with a) $I_P$=550 kA $q_{95}$=6.8 and b) $I_P$=500 kA $q_{95}$=6.8. Superimposed as circles and line are the q=m/3 rational surfaces of the discharge equilibrium.

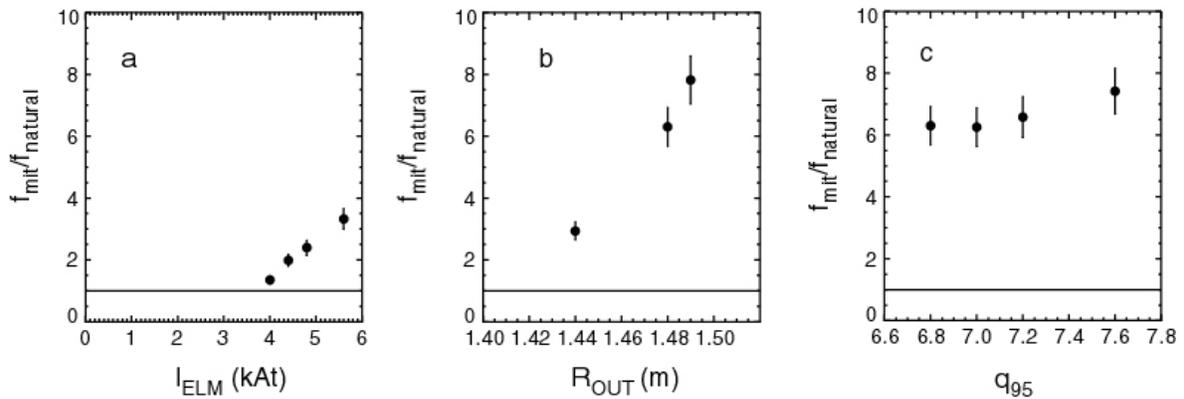

**Figure 9** The mitigated ELM frequency ($f_{mit}$) as a fraction of the natural ELM frequency ($f_{natural}$) obtained in scenario 6 plasmas as a function of a) ELM coil current ($I_{ELM}$), b) outer radius of the plasma ($R_{OUT}$) and c) edge safety factor ($q_{95}$).

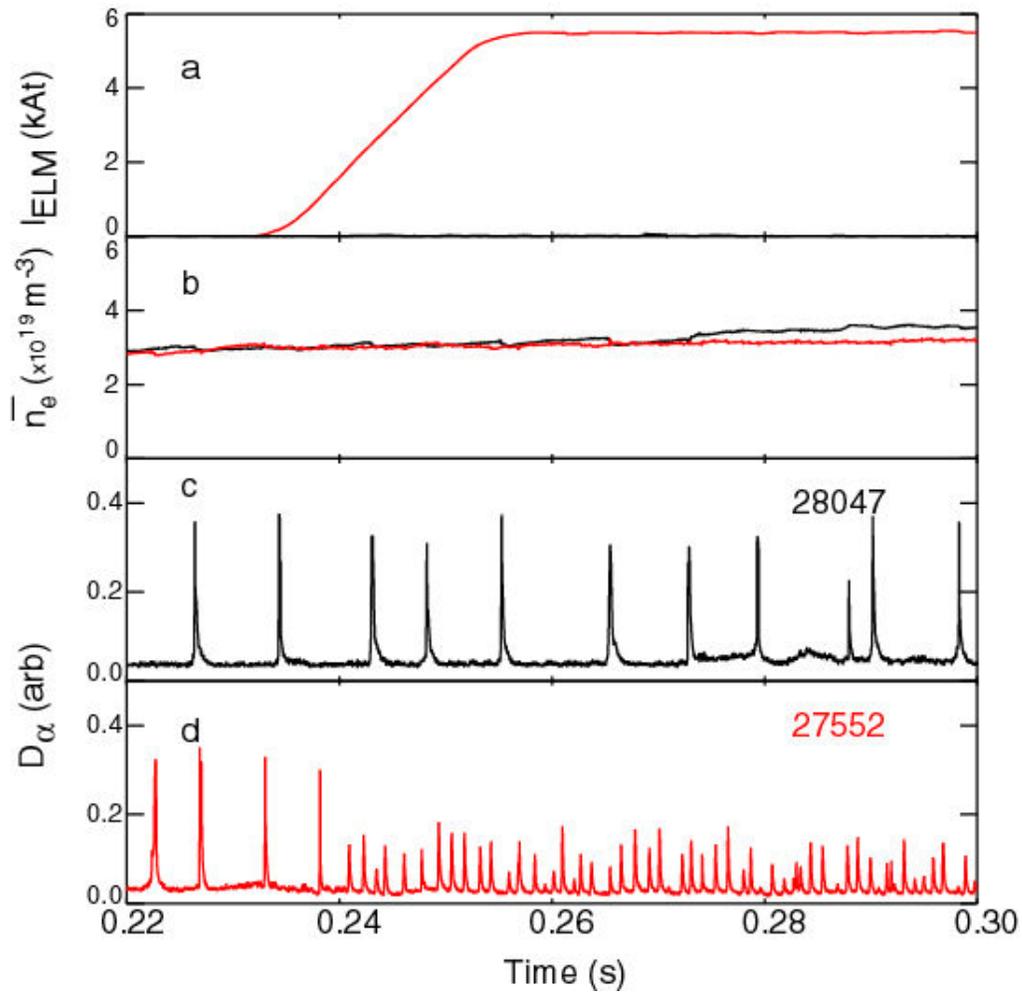

**Figure 10** Time traces for a scenario 6 shot with $I_P$=500 kA of a) coil current ($I_{ELM}$)  b) line average density ($\bar{n}_e$), and  $D_\alpha$ traces for c) shot  28047 without RMPs and d) shot 27552 with the RMPs in a even parity configuration.

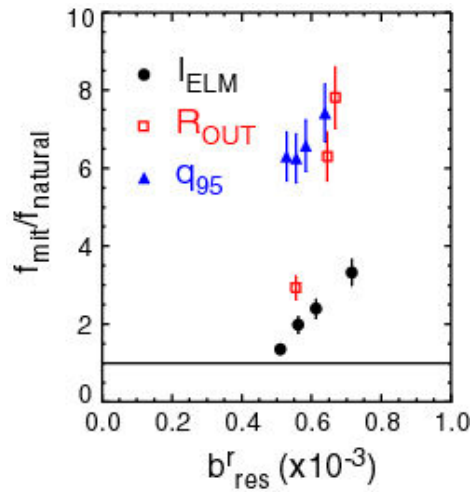

**Figure 11** The mitigated ELM frequency ($f_{mit}$) as a fraction of the natural ELM frequency ($f_{natural}$) obtained in scenario 6 plasmas as a function of the maximum resonant component of the applied field ($b^r_{res}$) calculated from vacuum modelling resulting from a scan in $I_{ELM}$ (circles), ROUT (squares) and q95 (triangles).

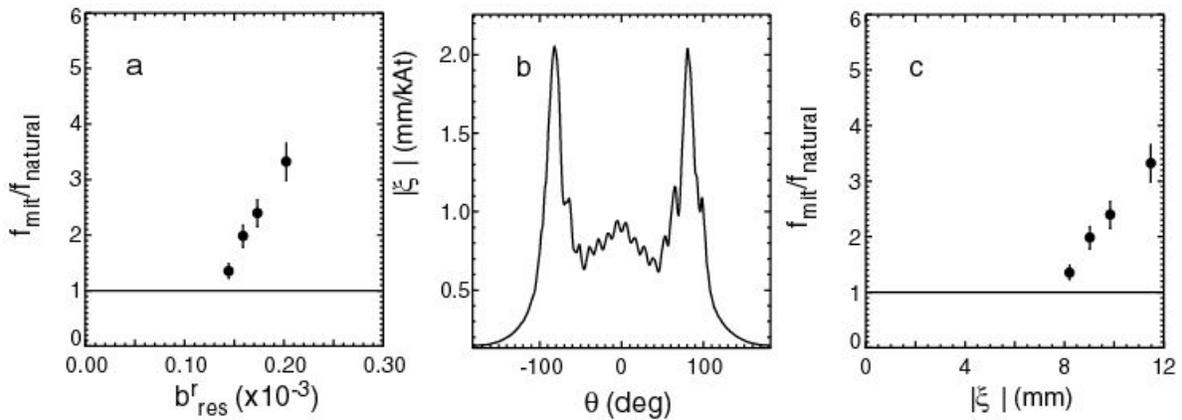

**Figure 12** Calculations of the plasma response computed by MARS-F for RMPs in an n=3 even parity configuration applied to a scenario 6 discharge: a) the mitigated ELM frequency ($f_{mit}$) as a fraction of the natural ELM frequency ($f_{natural}$) as a function of the maximum resonant component of the applied field ($b^r_{res}$), b) the amplitude of the normal displacement of the plasma surface ($\xi$) a function of poloidal angle ($\theta$) and c) $f_{mit}/f_{natural}$ as a function of $\xi$.

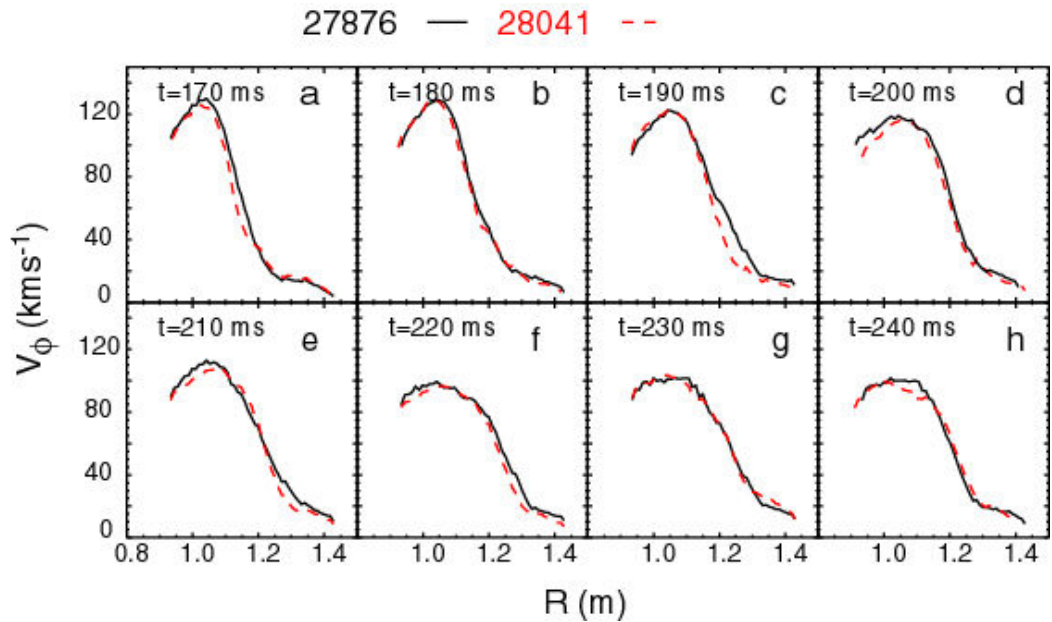

**Figure 13** The radial profile of the toroidal rotation velocity at 10 ms time intervals in the discharges shown in Figure 7 without (solid) and with (dashed) RMPs an n=3 even parity configuration.

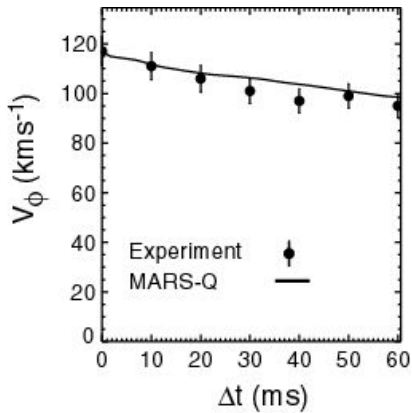

**Figure 14** The experimentally measured core toroidal rotation velocity (cirle) as a function of time after which the RMPs reached flat top (Δt) and the results from the MARS-Q code simulations (line).

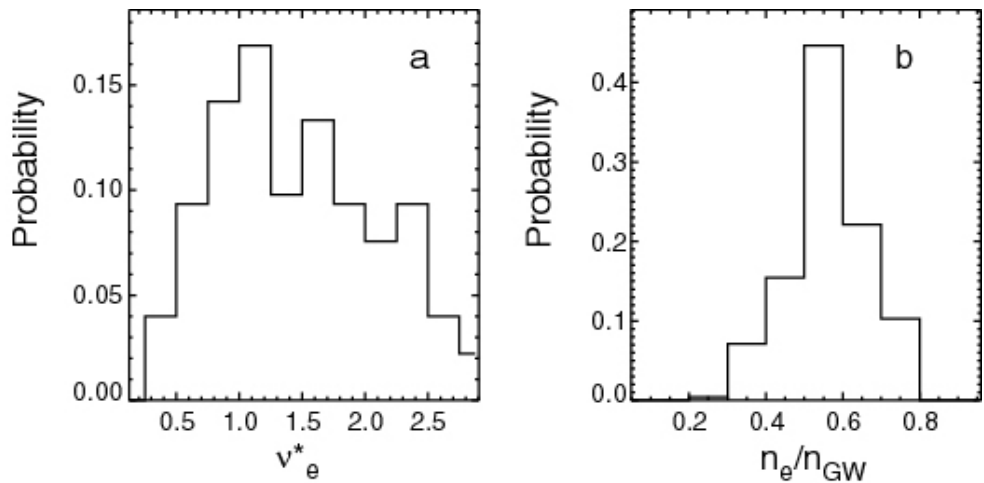

**Figure 15** Probability distribution of a) the edge pedestal edge collisionality ($v^*_e$) and b) the line averaged density as a fraction of the Greenwald density ($n_e/n_{GW}$).

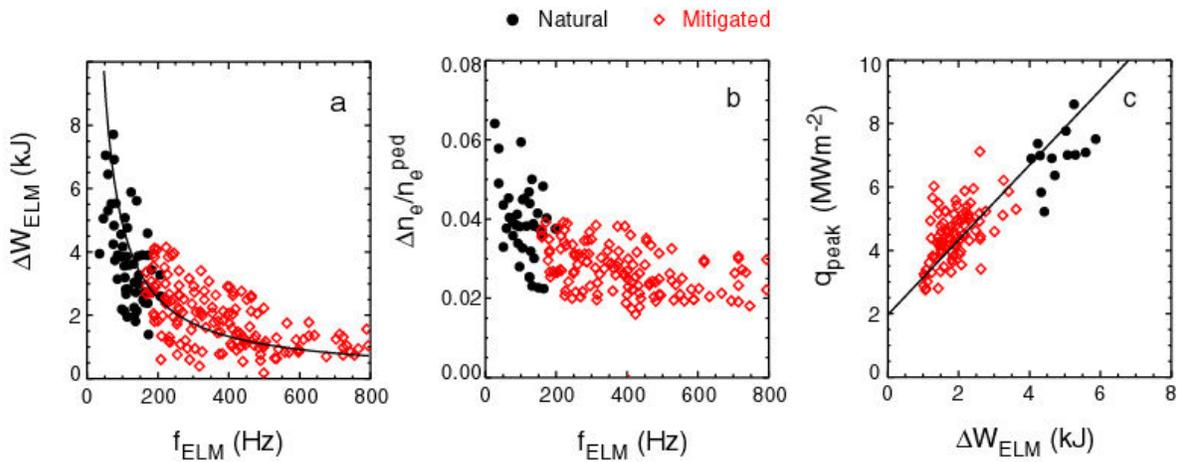

**Figure 16** a) ELM energy loss ($\Delta W_{ELM}$) as a function of ELM frequency ($f_{ELM}$) b) ELM particle loss expressed as a fraction of the pedestal density ($\Delta n_e/n_e^{ped}$) and c) maximum peak heat flux during an ELM at the low field side divertor as a function of $f_{ELM}$ for natural ($I_{ELM}=0$ kAt) and mitigated ELMs.

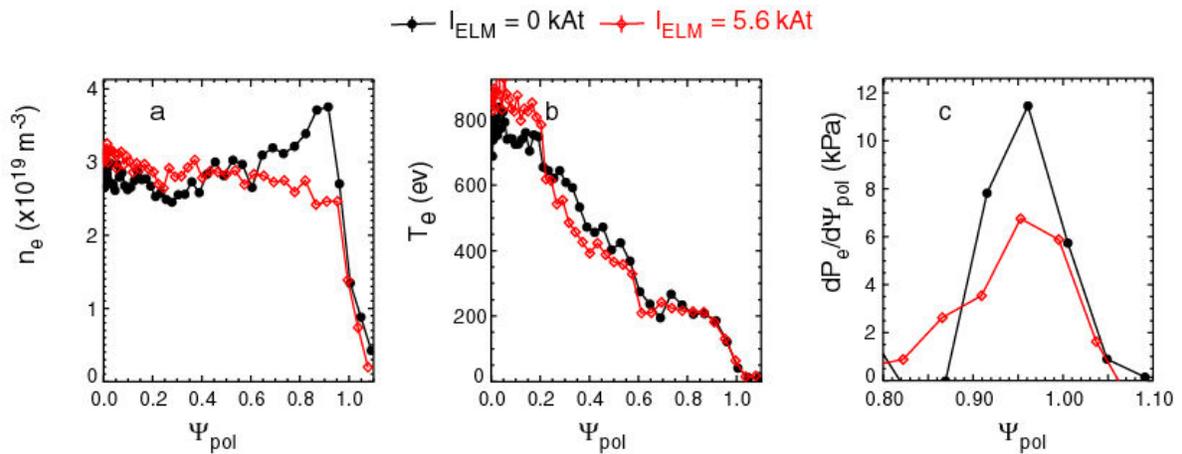

**Figure 17** Comparison of the profiles in a scenario 6 shot of electron a) density ($n_e$) , b) temperature ($T_e$) and c) pressure gradient ($dP_e/d\Psi_{pol}$) in normalised poloidal flux ($\Psi_{pol}$) space for shots with $I_{ELM} = 0$ kAt (closed circle) and $I_{ELM} = 5.6$ kAt (open diamonds) with the RMPs in an n=3 even parity configurations.

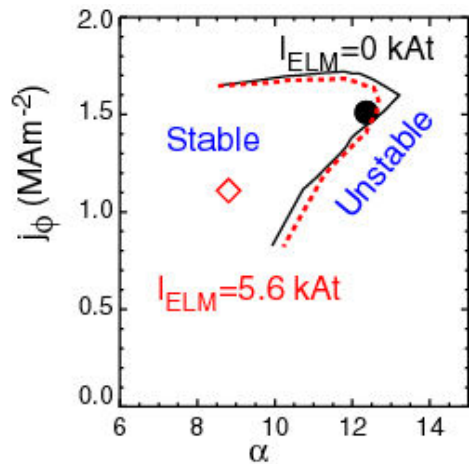

**Figure 18** Edge stability diagram plots of edge current density ($j_\phi$) versus normalised pressure gradient ($\alpha$) calculated for a shot with $I_{ELM} = 0$ kAt (solid line) and for a shot with $I_{ELM} = 5.6$ kAT in the RMPs in an n=3 configuration (dashed line). The solid circle and diamond represent the experimental points for the $I_{ELM} = 0$ and 5.6 kAt cases respectively.

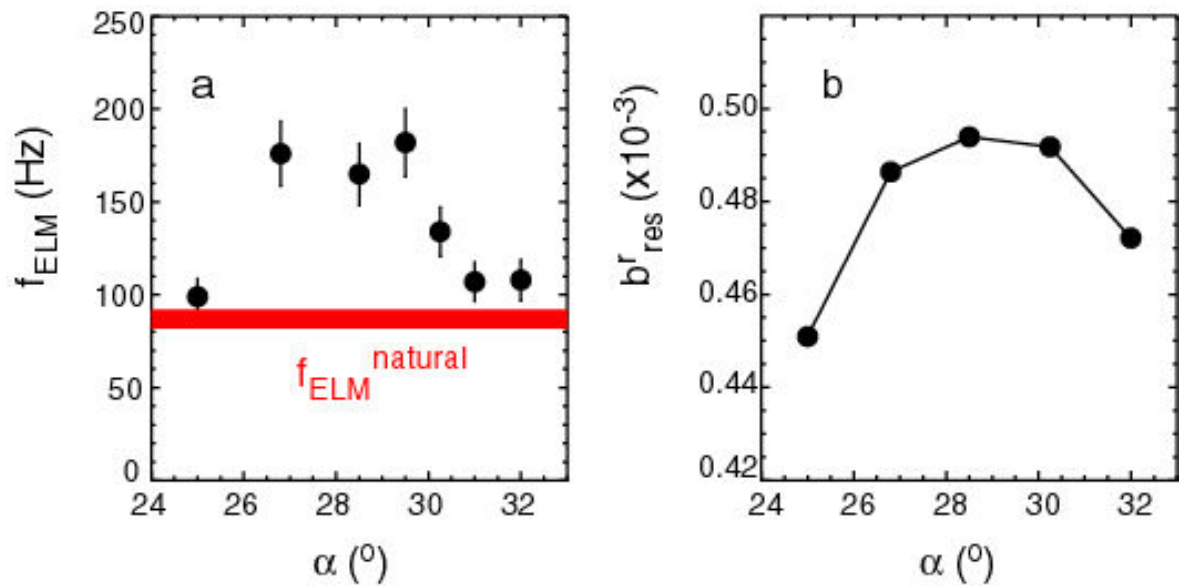

**Figure 19** a) ELM frequency as a function of the pitch angle (α) of the applied RMP field and b) the maximum resonant component of the applied field ($b^r_{res}$) calculated from vacuum modelling.